# Bifurcation Boundary Conditions for Switching DC-DC Converters Under Constant On-Time Control

Chung-Chieh Fang *

SUMMARY

Sampled-data analysis and harmonic balance analysis are applied to analyze switching DC-DC converters under constant on-time control. Design-oriented boundary conditions for the period-doubling bifurcation and the saddle-node bifurcation are derived. The required ramp slope to avoid the bifurcations and the assigned pole locations associated with the ramp are also derived. The derived boundary conditions are more general and accurate than those recently obtained. Those recently obtained boundary conditions become special cases under the general modeling approach presented in this paper. Different analyses give different perspectives on the system dynamics and complement each other. Under the sampled-data analysis, the boundary conditions are expressed in terms of signal slopes and the ramp slope. Under the harmonic balance analysis, the boundary conditions are expressed in terms of signal harmonics. The derived boundary conditions are useful for a designer to design a converter to avoid the occurrence of the period-doubling bifurcation and the saddle-node bifurcation.

**KEY WORDS:** DC-DC power conversion, constant on-time control, modeling

---

*C.-C Fang is with Advanced Analog Technology, 2F, No. 17, Industry E. 2nd Rd., Hsinchu 300, Taiwan, Tel: +886-3-5633125 ext 3612, Email: fangcc3@yahoo.com

# Contents

# 1 Introduction

Many efforts have been made in the past three decades to model the switching DC-DC converter under fixed frequency control [1, 2, 3, 4, 5]. Fewer efforts [6, 7, 8, 9] have been made to model the switching DC-DC converter under variable frequency control. Constant on-time control (COTC) is a type of variable frequency control. In two recent references [8, 9], converters under COTC, with the inductor current and the output voltage respectively as the feedback variable, are analyzed separately. In this paper, these two types of control schemes are analyzed in a unified model. The methodology proposed in this paper is based on the sampled-data analysis or the harmonic balance analysis, which provides an alternative way besides the tradition approaches to analyze the DC-DC converter under COTC. In [8, 9], approximate analysis is directly applied to analyze the converter. In this paper, *exact* sampled-data analysis or harmonic balance analysis is applied. Both the sampled-data analysis and the harmonic balance analysis produce the same results, but give different and complementary perspectives about the converter dynamics.

Two bifurcations commonly seen in DC-DC converters are analyzed. They are period-doubling bifurcation (PDB) and saddle-node bifurcation (SNB). The boundary conditions associated with the bifurcations are derived. Note that, here, the boundary condition means the critical condition in the converter *parameter space*, not about the well known critical eigenvalues in the *complex plane*. The boundary conditions define the bifurcation boundaries in the parameter space to separate stable and unstable regions. The methodology proposed here can be applied to general DC-DC converters, and the buck converter is used as an example throughout the paper. It will be shown that the boundary conditions obtained in [7, 8, 9] become special cases in terms of the general models used in this paper. Although the sampled-data analysis has been applied to obtain similar boundary conditions in [7], the approach applied here is different. In [7], an *approximate* sampled-data model is directly applied, whereas here an *exact* sampled-data model is directly applied and further approximations are then applied later to preserve the accuracy.

The remainder of the paper is organized as follows. In Section 2, the operation of COTC is reviewed. In Section 3, exact sampled-data analysis is applied. In Sections 4 and 5, PDB and SNB are respectively analyzed based on the sampled-data analysis. In Sections 6 and 7, PDB and SNB are respectively analyzed based on the harmonic balance analysis. Conclusions are collected in Section 8.

# 2 Operation of Constant On-Time Control (COTC)

The operation of a switching DC-DC converter under COTC can be described *exactly* by a unified block diagram model [10, 11] shown in Fig. 1. Denote the control signal as $v_c$ which controls the inductor current $i_L$ or the output voltage $v_o$ as discussed below. Denote the source voltage as $v_s$. In the model, $A_1, A_2 \in \mathbf{R}^{N \times N}$, $B_1, B_2 \in \mathbf{R}^{N \times 2}$, $C, E_1, E_2 \in \mathbf{R}^{1 \times N}$, and $D \in \mathbf{R}^{1 \times 2}$ are constant matrices, where $N$ is the system dimension. In the $n$-th cycle, the dynamics is switched between two stages, $S_1$ and $S_2$. Switching from $S_1$ to $S_2$ occurs at $t = \sum_{i=1}^{n-1} T_i + d$ (where $d$ is fixed in COTC). Switching from $S_2$ to $S_1$ occurs at $t = \sum_{i=1}^{n} T_i$ (where $T_i$ varies in each cycle) when the ramp signal $h(t)$ intersects with the signal $y := Cx + Du \in \mathbf{R}$.

In this paper, two COTC schemes are analyzed. The first scheme is the valley current control (a variation of current mode control but with COTC, see Fig. 2 for a buck converter). One has $y = R_i i_L - v_c$, where $R_i$ is the sensing resistance. At the switching instants, one has $y(t) = h(t)$, or equivalently, $R_i i_L = v_c + h(t)$. The converter changes from $S_2$ to $S_1$ when $R_i i_L$ drops below $v_c + h(t)$. The second scheme is the valley voltage control (a variation of V$^2$ control but with COTC, see Fig. 3 for a buck converter). One has $y = v_o - v_c$. The converter changes from $S_2$ to $S_1$ when $v_o$ drops below $v_c + h(t)$. In this paper, these two COTC schemes are denoted as C-COTC and V-COTC, respectively. In C-COTC, the inductor *current* is the feedback signal, and it is also called current mode COTC in [8]. In V-COTC, the output *voltage* is the feedback signal, and it is also called $V^2$ COTC in [9].

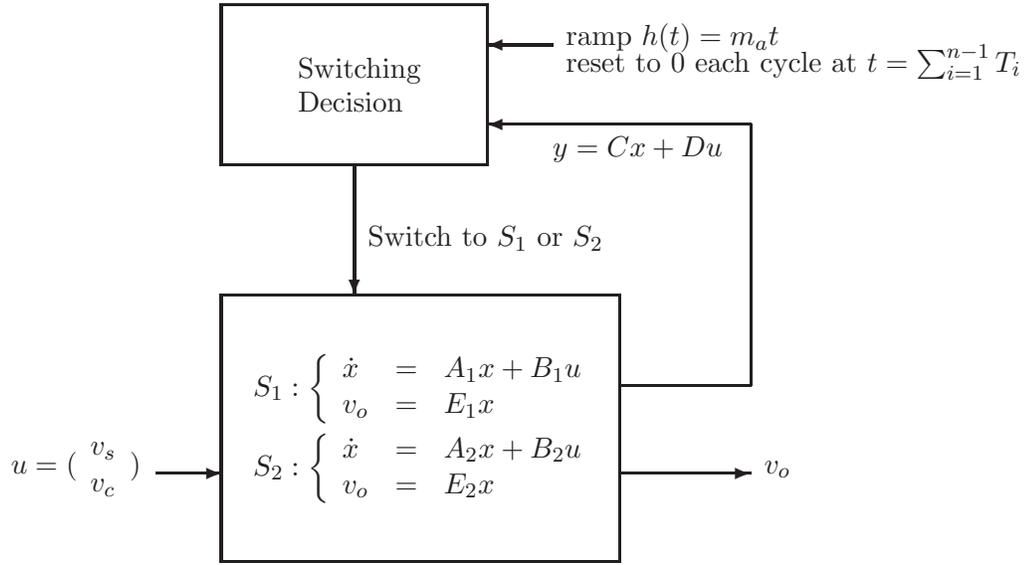

Figure 1: Block diagram model for switching converter under variable frequency control.

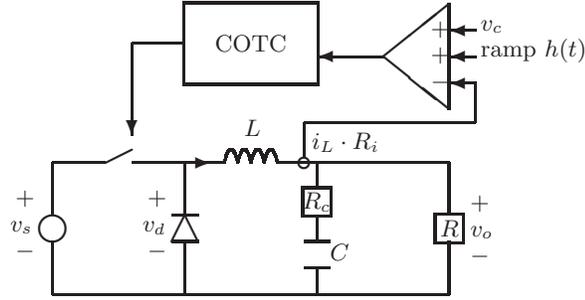

Figure 2: A buck converter under C-COTC.

Denote the ramp slope as $m_a$. In the $n$-th cycle, denote the ramp amplitude as $V_{hn} = m_a T_n$. The short notation $V_{hn}$, instead of $V_{h,n}$, is used for brevity. This applies to other variables. Since $T_n$ varies for each cycle, the ramp amplitude $V_h$ also varies for each cycle. Let the steady-state of the period be $T$. In steady state, $V_{hn} = V_h = m_a T$. Denote the switching frequency as $f_s := 1/T$ and let $\omega_s := 2\pi f_s$. In COTC, the switching occurs at $t = \sum_{i=1}^{n-1} T_i + d$ where $d$ is fixed, and the period $T_n$ is used as a control variable. This is different from the fixed frequency control, where the switching occurs at $t = nT + d_n$ where $d_n$ is controlled, and the switching period $T$ is fixed.

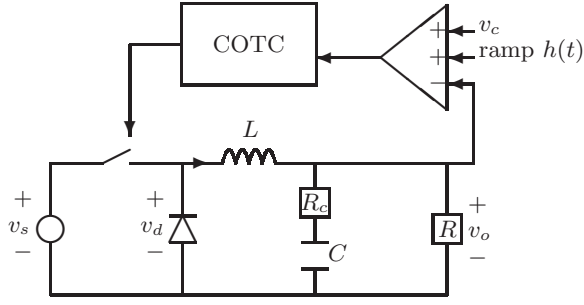

Figure 3: A buck converter under V-COTC.

# 3 Exact Sampled-Data Dynamics of a General Converter

## 3.1 Large-Signal Analysis

In the fixed frequency control, the states are sampled at $t = nT$ with a fixed period. In COTC, the states are sampled at $t = \sum_{i=1}^{n-1} T_i$ with a variable period. Let $u_n = (v_{sn}, v_{cn})'$. Similar to the analysis in [11], the large-signal nonlinear constrained dynamics, mapping from $x_n := x(\sum_{i=1}^{n-1} T_i)$ to $x_{n+1}$, is

$$\begin{aligned} x_{n+1} &= f(x_n, u_n, T_n) \\ &= e^{A_2(T_n-d)}(e^{A_1 d}x_n + \int_0^d e^{A_1\sigma}d\sigma B_1 u_n) + \int_d^{T_n} e^{A_2(T-\sigma)}d\sigma B_2 u_n \end{aligned} \quad (1)$$

$$\begin{aligned} g(x_n, u_n, T_n) &= Cf(x_n, u_n, T_n) + Du_n - m_a T_n \\ &= 0 \end{aligned} \quad (2)$$

where the constraint $g(x_n, u_n, T_n) = 0$ determines the switching instant $T_n$. In the $n$-th period, once the variable period $T_n$ is determined, $x_n$ is mapped to $x_{n+1}$ through the function $f$.

## 3.2 Steady-State Analysis

The periodic solution $x^0(t)$ of the system in Fig. 1 corresponds to a fixed point $x^0(0)$ in the sampled-data dynamics. Let $\dot{x}^0(0^-) = A_2 x^0(0) + B_2 u$ denote the time derivative of $x^0(t)$ at $t = 0^-$. Let $y^0(t) = Cx^0(t) + Du$. In steady state, $\dot{y}^0(t) = C\dot{x}^0(t)$. Let the steady-state duty cycle be $D$, then $d = DT$. Confusion of notations for the capacitance $C$ and the duty cycle $D$ with the matrices $C$ and $D$ can be avoided from the context.

In steady state,

$$x^0(d) = e^{A_1 d}x^0(0) + \int_0^d e^{A_1\sigma}d\sigma B_1 u \quad (3)$$

$$x^0(0) = e^{A_2(T-d)}x^0(d) + \int_0^{T-d} e^{A_2\sigma}d\sigma B_2 u \quad (4)$$

From (3) and (4), one has

$$x^0(0) = (I - e^{A_2(T-d)}e^{A_1 d})^{-1}(e^{A_2(T-d)}\int_0^d e^{A_1\sigma}d\sigma B_1 u + \int_0^{T-d} e^{A_2\sigma}d\sigma B_2 u) \quad (5)$$

Let $B_1 := [B_{11}, B_{12}]$ and $B_2 := [B_{21}, B_{22}]$ to expand the matrices into two columns. The COTC *buck* converter generally has $A_1 = A_2$, $B_{21} = 0$, and $B_{12} = B_{22}$. Then, from (3) and (4), one has

$$x^0(0) = (I - e^{A_1 T})^{-1} A_1^{-1} (e^{A_1 T} - e^{A_1 (T-d)}) B_{11} v_s - A_1^{-1} B_{12} v_c \tag{6}$$

Generally the controller may include an integrator (with a pole at zero), making $A_1$ or $I - e^{A_2(T-d)} e^{A_1 d}$ non-invertible. In that case, the pole at zero can be replaced by a very small number $\delta$, then $A_1$ or $I - e^{A_2(T-d)} e^{A_1 d}$ are invertible. Therefore, the invertibility of $A_1$ or $I - e^{A_2(T-d)} e^{A_1 d}$ is not critical and can be resolved. This statement about invertibility of a matrix is not repeated later.

The boost converter generally has $B_1 = B_2$, then

$$x^0(0) = (I - e^{A_2(T-d)} e^{A_1 d})^{-1} (e^{A_2(T-d)} \int_0^d e^{A_1 \sigma} d\sigma + \int_0^{T-d} e^{A_2 \sigma} d\sigma) B_1 u \tag{7}$$

### 3.3 Small-Signal Analysis

Using a hat ˆ to denote small perturbations (e.g., $\hat{x}_n = x_n - x^0(0)$). From [2, 10, 11, 12], the linearized sampled-data dynamics is

$$\hat{x}_{n+1} = \Phi \hat{x}_n + \Gamma \hat{u}_n = \Phi \hat{x}_n + \Gamma_1 \hat{v}_{sn} + \Gamma_2 \hat{v}_{cn} \tag{8}$$

where $\Phi \in \mathbf{R}^{N \times N}$ and $\Gamma = [\Gamma_1, \Gamma_2] \in \mathbf{R}^{N \times 2}$ (expanded into two columns) are

$$\begin{aligned}
\Phi &= \left. \frac{\partial f}{\partial x_n} - \frac{\partial f}{\partial T_n} \left( \frac{\partial g}{\partial T_n} \right)^{-1} \frac{\partial g}{\partial x_n} \right|_{(x_n, u_n, T_n) = (x^0(0), u, T)} \\
&= (I - \frac{\dot{x}^0(0^-) C}{C \dot{x}^0(0^-) - m_a}) e^{A_2(T-d)} e^{A_1 d}
\end{aligned} \tag{9}$$

$$\begin{aligned}
\Gamma &= \left. \frac{\partial f}{\partial u_n} - \frac{\partial f}{\partial T_n} \left( \frac{\partial g}{\partial T_n} \right)^{-1} \frac{\partial g}{\partial u_n} \right|_{(x_n, u_n, T_n) = (x^0(0), u, T)} \\
&= (I - \frac{\dot{x}^0(0^-) C}{C \dot{x}^0(0^-) - m_a})(e^{A_2(T-d)} \int_0^d e^{A_1 \sigma} d\sigma B_1 + \int_0^{T-d} e^{A_2 \sigma} d\sigma B_2) - \frac{\dot{x}^0(0^-) D}{C \dot{x}^0(0^-) - m_a}
\end{aligned} \tag{10}$$

Local orbital stability of the converter is determined by the eigenvalues of $\Phi$, which are also the sampled-data poles of the converter. The periodic solution $x^0(t)$ is asymptotically orbitally stable if all of the eigenvalues of $\Phi$ are inside the unit circle of the complex plane.

The PDB occurs when one eigenvalue of $\Phi$ is $-1$, and $\det[I + \Phi] = 0$. The SNB occurs when one eigenvalue of $\Phi$ is 1, and $\det[I - \Phi] = 0$. Since other bifurcations [13], such as pitchfork or transcritical bifurcations, also have one eigenvalue of $\Phi$ at 1, the same analysis can be applied to analyze these bifurcations, omitted to save space. All slope-based boundary conditions derived in this paper are based on the closed form expression of (9).

### 3.4 Control-to-Output and Audio-Susceptibility Transfer Functions

In (10), with the matrices $B_1$, $B_2$, and $D$ replaced by their first columns, then $\Gamma$ becomes $\Gamma_1$. Similarly, with the matrices $B_1$, $B_2$, and $D$ replaced by their second columns, then $\Gamma$ becomes $\Gamma_2$. Since the matrix $D = [0, -1]$ for C-COTC and V-COTC and the second columns of the matrices

$B_1$ and $B_2$ are generally zeros, the matrices $\Gamma_1$ and $\Gamma_2$ can be further simplified as

$$\Gamma_1 = (I - \frac{\dot{x}^0(0^-)C}{C\dot{x}^0(0^-) - m_a})(e^{A_2(T-d)} \int_0^d e^{A_1\sigma} d\sigma B_{11} + \int_0^{T-d} e^{A_2\sigma} d\sigma B_{21}) \tag{11}$$

$$\Gamma_2 = \frac{\dot{x}^0(0^-)}{C\dot{x}^0(0^-) - m_a} \tag{12}$$

Since the output voltage may be discontinuous (as in the boost converter), let $E := (E_1 + E_2)/2$ so that $Ex$ is the average output voltage. From (8), the control-to-output transfer function is

$$T_{oc}(z) = \frac{\hat{v}_o(z)}{\hat{v}_c(z)} = E(zI - \Phi)^{-1}\Gamma_2 \tag{13}$$

The DC gain is $T_{oc}(e^{j0}) = T_{oc}(1) = E(I - \Phi)^{-1}\Gamma_2$. Given a transfer function in the $z$ domain, say $T(z)$, its effective frequency response [14, p. 93] is $T(e^{j\omega T})$, which is valid up to half the switching frequency.

Also from (8), the audio-susceptibility is

$$T_{os}(z) = \frac{\hat{v}_o(z)}{\hat{v}_s(z)} = E(zI - \Phi)^{-1}\Gamma_1 \tag{14}$$

Other transfer functions, such as output impedance, can be similarly derived as in [10].

### 3.5 General Slope-Based Boundary Conditions for PDB and SNB

**Theorem 1** *Suppose that $\lambda$ is not an eigenvalue of $e^{A_2(T-d)}e^{A_1d}$. Then $\lambda$ is an eigenvalue of $\Phi$ if and only if*

$$C(I - \lambda^{-1}e^{A_1d}e^{A_2(T-d)})^{-1}\dot{x}^0(0^-) = m_a \tag{15}$$

*Proof:* Suppose $\lambda$ is not an eigenvalue of $e^{A_2(T-d)}e^{A_1d}$, then

$$\begin{aligned}\det[\lambda I - \Phi] &= \det[\lambda I - e^{A_2(T-d)}e^{A_1d}] \cdot \\ &\quad \det[I + (\lambda I - e^{A_2(T-d)}e^{A_1d})^{-1}(\frac{\dot{x}^0(0^-)C}{C\dot{x}^0(0^-) - m_a})e^{A_2(T-d)}e^{A_1d}] \\ &= \det[\lambda I - e^{A_2(T-d)}e^{A_1d}] \cdot \\ &\quad (1 + Ce^{A_2(T-d)}e^{A_1d}(\lambda I - e^{A_2(T-d)}e^{A_1d})^{-1}(\frac{\dot{x}^0(0^-)}{C\dot{x}^0(0^-) - m_a}))\end{aligned}$$

Therefore, $\lambda$ is an eigenvalue of $\Phi$ if and only if

$$1 + Ce^{A_2(T-d)}e^{A_1d}(\lambda I - e^{A_2(T-d)}e^{A_1d})^{-1}(\frac{\dot{x}^0(0^-)}{C\dot{x}^0(0^-) - m_a}) = 0$$

which can be rearranged as

$$C\dot{x}^0(0^-) + C(\lambda e^{-A_2(T-d)}e^{-A_1d} - I)^{-1}\dot{x}^0(0^-) = m_a \tag{16}$$

leading to (15) based on the matrix equality $I + (\lambda e^{-A_2(T-d)}e^{-A_1d} - I)^{-1} = (I - \lambda^{-1}e^{A_1d}e^{A_2(T-d)})^{-1}$. $\square$

Note that the condition (15) is applicable to *general* switching converters of any system dimension. Also note that in (15), the left side is related to the ripple slope $\dot{x}^0(0^-)$, and the right side is the ramp slope $m_a$. As in the popular slope-based boundary condition for the subharmonic oscillation in the current-mode control, the condition (15) is also *slope*-based. For designation purpose, the left side of (15) is called an "S plot", $S(\lambda, D)$, as a function of $\lambda$ and $D = d/T$, for example. The S plot can be a function of other variables, such as $T$, and the power stage parameters, such as $R$, $L$, and $C$.

**Corollary 1**
*(i) If the system parameters correspond to an occurrence of period-doubling bifurcation ($\lambda = -1$), then*

$$S(-1, D) = C(I + e^{A_1 TD} e^{A_2 T(1-D)})^{-1} \dot{x}^0(0^-) = m_a \qquad (17)$$

*(ii) If the system parameters correspond to an occurrence of saddle-node bifurcation ($\lambda = 1$), then*

$$S(1, D) = C(I - e^{A_1 TD} e^{A_2 T(1-D)})^{-1} \dot{x}^0(0^-) = m_a \qquad (18)$$

*(iii) Without the ramp ($m_a = 0$), then $\det[\Phi] = 0$, and $\Phi$ has an eigenvalue of 0. Equivalently, $S(0, D) = 0$.*

*Proof:* Proof for *(i)* or *(ii)* is directly from Theorem 1. Proof for *(iii)* is as follows.
From (9) with $m_a = 0$, then $\det[\Phi] = \det[I - \dot{x}^0(0^-)C/(C\dot{x}^0(0^-))] \det[e^{A_2(T-d)} e^{A_1 d}] = \det[1 - C\dot{x}^0(0^-)/(C\dot{x}^0(0^-))] \det[e^{A_2(T-d)} e^{A_1 d}] = 0$. Another proof for *(iii)* is based on the fact that $\lambda = 0$ is a root of (16) with $m_a = 0$. □

A sampled-data pole at 0 generally causes deadbeat effect [15] where transient perturbation correction occurs within $N$ switching periods [7]. A sampled-data pole at 0 also generally causes the continuous-time dynamics to have a quality factor $Q = 2/\pi$ [3].

## 3.6 Pole Locus and Pole Assignment

In the traditional root locus plot, the root locus is a function of the feedback gain. In COTC, the ramp slope $m_a$ determined by (15) can be expressed as a function of the real pole $\lambda$. Without the ramp ($m_a = 0$), one pole is always zero. With the ramp, the poles are shifted according to (15). For designation purpose, the S plot as a function of the *real* pole $\lambda$ *alone*, is called a pole locus plot. Given a desired real pole location, one can see the required ramp slope for pole assignment based on the pole locus plot (15).

In summary, the S plot is useful for design purpose to avoid PDB and SNB instabilities, and to assign the real poles. First, the S plot $S(-1, D)$ can predict the occurrence of PDB and the required ramp slope to avoid PDB. For example, given a ramp slope $m_a$, the intersection of the S plot and the horizontal line at $m_a$ shows the unstable operating range of the duty cycle $D$ if $S(\lambda, D) > m_a$. To avoid PDB, one need to increase the ramp slope $m_a$ such that $S(\lambda, D) < m_a$ for the operating range of $D$. Similarly, the S plot $S(1, D)$ can predict the occurrence of SNB and the required ramp slope to avoid SNB.

Second, given a duty cycle $D$, a special case of the S plot, $S(\lambda, D)$, as a function of $\lambda$ alone (also called pole locus plot), can predict the *real* pole locations and the required ramp slope for pole assignment. For example, the intersection of the S plot and the horizontal line at $m_a$ shows the real pole location. As the ramp slope varies, the pole location also varies. One can see the required ramp slope to assign the real pole to a particular location. If the S plot does not intersect with the horizontal line at $m_a$, there are no real poles. Complex poles may exist in that situation.

## 3.7 Approximate Pole Locus Plot

The *exact* pole locus plot $S(\lambda, D)$, the left side of (15), can be expressed in an approximate form. Based on the assumption that $RC$ and $\sqrt{LC}$ are much larger than $T$, matrix approximations such as $e^{A_1 T} \approx I + A_1 T$ and $(I + A_1 T)^{-1} \approx I - A_1 T$ can be applied. Then, (15) leads to

$$S(\lambda, D) \approx \frac{\lambda C}{\lambda - 1}(I + \frac{A_1 d + A_2(T-d)}{\lambda - 1})\dot{x}^0(0^-) = m_a \tag{19}$$

Without the ramp ($m_a = 0$), (19) leads to

$$\lambda = 1 - \frac{C(A_1 d + A_2(T-d))\dot{x}^0(0^-)}{C\dot{x}^0(0^-)} \tag{20}$$

For the buck converter with $A_1 = A_2$ and $B_2 = 0_{2 \times 2}$, then $\dot{x}^0(0^-) = A_1 x^0(0)$ and (20) becomes

$$\lambda = 1 - \frac{CA_1^2 T x^0(0)}{CA_1 x^0(0)} \tag{21}$$

where the *exact* expression of $x^0(0)$ can be obtained from (5). Therefore, without the ramp ($m_a = 0$), two poles of the buck converter are 0 and (21). Let the state be $x = (i_L, v_C)'$, where $i_L$ is the inductor current and $v_C$ is the capacitor voltage. Then, for the buck converter, $x^0(0)$ can be *approximated* as $[v_s D/R - v_s D(1-D)T/2L, v_s D]'$ based on the traditional average analysis [15].

## 4 Sampled-Data Analysis of PDB in the Buck Converter

The analysis above is for the general DC-DC converter. The general analysis is applied below to the buck converter. Many critical parameters (such as the ramp slope and the constant on-time) at the stability/instability boundary can be derived based on the boundary conditions.

**Using the boundary condition to determine the minimum ramp slope.**
The buck converter generally has $A_1 = A_2$, $B_{21} = 0$, and $B_{12} = B_{22}$. Using (6), the PDB boundary condition (17) becomes

$$S(-1, D) = C(I - e^{2A_1 T})^{-1}(e^{A_1 T} - e^{A_1 T(1-D)})B_{11} v_s = m_a \tag{22}$$

Generally, with larger $m_a$ or smaller $v_s$, the converter is stable. Therefore, the converter is stable if

$$S(-1, D) < m_a \tag{23}$$

Based on the assumption that $RC$ and $\sqrt{LC}$ are much larger than $T$, matrix approximations such as $e^{A_1 T} \approx I + A_1 T + A_1^2 T^2/2$ and $(I + A_1 T)^{-1} \approx I - A_1 T$ can be applied. Then, the boundary condition (23) leads to

$$S(-1, D) \approx (-\frac{D}{2}CB_{11} + (\frac{D^2}{4})CA_1 B_{11} T + (\frac{D - D^3}{12})CA_1^2 B_{11} T^2)v_s < m_a \tag{24}$$

For $CA_1 B_{11} \gg CA_1^2 B_{11} T$ (or equivalently, $RC \gg T$, in V-COTC, for example), (24) leads to

$$(-\frac{D}{2}CB_{11} + (\frac{D^2}{4})CA_1 B_{11} T)v_s < m_a \tag{25}$$

or equivalently, using the fact that $d = DT$,

$$\frac{Dv_s}{2}(\frac{d}{2}CA_1B_{11} - CB_{11}) < m_a \qquad (26)$$

which shows the required ramp slope $m_a$ to stabilize the converter to avoid PDB. The required ramp slope is linearly proportional to $D$. Therefore, to operate at a higher duty cycle, a larger ramp slope is required. Also note that $v_o \approx v_s D$ for the buck converter. Based on (26) and given a fixed $v_o$, the required ramp slope to avoid subharmonic oscillation is the same, independent of $v_s$. This indicates that COTC has very good *line* regulation to avoid the subharmonic oscillation. However, note that the condition (26) is approximate under the assumption that $T$ is small, the exact condition is (22). In COTC, $d$ is fixed and $T$ may vary to a large degree. For a large steady-state value of $T$, (24) is more accurate than (25). However, (24) is complex. One can use (25) for simplicity at the expense of accuracy.

**Using the boundary condition to determine the maximum constant on-time.**
The boundary condition (26) can be expressed in terms of the maximum constant on-time $d$,

$$\frac{d}{2} < \frac{CB_{11} + \frac{2m_a}{Dv_s}}{CA_1B_{11}} \quad \text{if } CA_1B_{11} > 0 \qquad (27)$$

The inequality sign is reversed if $CA_1B_{11} < 0$.

Added with external ramps, the two control schemes, V-COTC and C-COTC, with or without external ramps added, are analyzed as follows. For V-COTC, one can also use the inductor current as the ramp instead of the external ramp $h(t)$.

## 4.1  V-COTC Added With an External Ramp ($m_a \neq 0$)

In V-COTC, $y = v_o - v_c$. let the load be $R$, the inductance be $L$, the capacitance be $C$, and the equivalent series resistance (ESR) be $R_c$. Let $\rho = R/(R + R_c)$. For $R_c = 0$, $\rho = 1$. Then,

$$A_1 = A_2 = \rho \begin{bmatrix} \frac{-R_c}{L} & \frac{-1}{L} \\ \frac{1}{C} & \frac{-1}{RC} \end{bmatrix} \qquad (28)$$

$$B_{11} = \begin{bmatrix} \frac{1}{L} \\ 0 \end{bmatrix} \qquad (29)$$

$$E_1 = E_2 = \rho \begin{bmatrix} R_c & 1 \end{bmatrix} \qquad (30)$$

$$\begin{array}{ll} C = \rho[R_c, 1] & D = [0, -1] \\ CB_{11} = \frac{\rho R_c}{L} & CA_1B_{11} = \frac{\rho^2}{LC}(1 - \frac{R_c^2 C}{L}) \end{array} \qquad (31)$$

The boundary conditions (26) and (27) become, respectively,

$$m_a > \frac{Dv_s\rho^2}{2LC}(\frac{d}{2}(1 - \frac{R_c^2 C}{L}) - \frac{R_c C}{\rho}) \qquad (32)$$

$$\frac{d}{2} < \frac{\frac{\rho R_c}{L} + \frac{2m_a}{Dv_s}}{\frac{\rho^2}{LC}(1 - \frac{R_c^2 C}{L})} \qquad (33)$$

which are further simplified to (if $R_c^2 C \ll L$ and $R_c \ll R$)

$$m_a > \frac{Dv_s}{2LC}(\frac{d}{2} - R_c C) \qquad (34)$$

$$\frac{d}{2} < R_c C + \frac{2m_a LC}{Dv_s} \qquad (35)$$

Generally the required ramp slope condition is normalized by $s_f := R_c v_s D/L$, the off-time inductor current slope multiplied by $R_c$. Note that $s_f$ is also the output voltage ripple slope contributed by the inductor current. The required ramp slope (32) becomes

$$\frac{m_a}{s_f} > \frac{\rho^2}{2}\left(\frac{d}{2R_cC}\left(1 - \frac{R_c^2 C}{L}\right) - \frac{1}{\rho}\right) \tag{36}$$

For $R_c^2 C \ll L$ and $R_c \ll R$, (36) is further simplified to

$$\frac{m_a}{s_f} > \frac{1}{2}\left(\frac{d}{2R_cC} - 1\right) \tag{37}$$

which agrees with [9, Eq. 21]. Note that (37) is a special case of the general condition (17) (applicable to all kinds of converters). Also note that if the conditions $R_c^2 C \ll L$ and $R_c \ll R$ are not met, (36) will be shown to be more accurate than the boundary condition (37) as obtained in [9].

## 4.2 V-COTC Without an External Ramp ($m_a = 0$)

Without the ramp ($m_a = 0$), the boundary condition (33) becomes

$$\frac{d}{2} < \frac{R_c C}{\rho(1 - \frac{R_c^2 C}{L})} \tag{38}$$

For $R_c^2 C \ll L$ and $R_c \ll R$, (38) is further simplified to

$$\frac{d}{2} < R_c C \tag{39}$$

which agrees with [7, 9]. Note that if the conditions $R_c^2 C \ll L$ and $R_c \ll R$ are not met, (38) will be more accurate than the boundary condition (39) as obtained in [7, 9].

**Pole Locus**
The pole location can be also expressed in a closed form. With some approximations for $R_c T \ll L$, (21) leads to

$$\lambda = 1 + \frac{2\rho T}{2R_c C + T - d}\left(\frac{T - d}{2RC} - 1\right) \tag{40}$$

By setting $\lambda > -1$ in (40), the maximum of on-time to avoid subharmonic oscillation is

$$\frac{d}{2} < \frac{R_c C + \frac{T^2}{4(R+R_c)C}}{1 + \frac{T}{2(R+R_c)C}} \tag{41}$$

Based on some examples, (41) is more accurate than (38) and (39). Note that (41) depends on $T$, whereas (38) and (39) are independent of $T$. For $T \ll (R + R_c)C$, (41) becomes (39).

For $T - d \ll RC$ and $R_c \ll R$, (40) is further simplified to

$$\lambda = 1 - \frac{2T}{2R_c C + T - d} \tag{42}$$

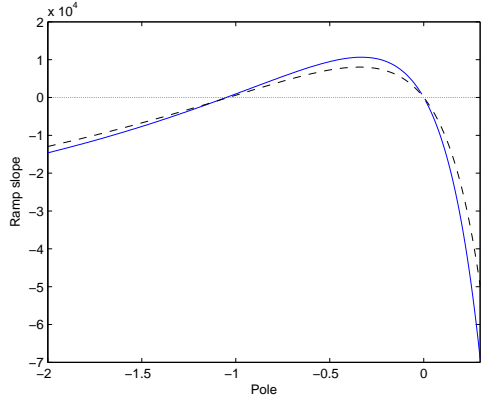

Figure 4: The pole locus plot $S(\lambda, 0.4)$ shows the required ramp slope as a function of the pole $\lambda$. Its intersection with the horizontal line at $m_a = 0$ shows the poles at 0 and -1.1, (solid line for the exact expression (15) and dashed line for the approximate expression (19)).

which agrees with [7, p. 2679]. Note that if the conditions $T - d \ll RC$ and $R_c \ll R$ are not met, (40) will be more accurate than the pole location (42) as obtained in [7]. This sampled-data pole can be mapped to an equivalent continuous-time pole,

$$\frac{2}{2R_cC + T - d} = \frac{1}{R_cC + \frac{T(1-D)}{2}} \tag{43}$$

which is close to the pole $1/R_cC$ derived in [9, 16] if $R_cC \gg T$.

**Example 1.** (*Without compensating ramp.*) Consider a V-COTC buck converter from [7] with the following parameters: $v_s = 5$ V, $T = 3$ μs, $d = 1.2$ μs, $R = 0.5$ Ω, $R_c = 20$ mΩ, $L = 2$ μH, and $C = 20$ μF. Here, $D = d/T = 0.4$. With these parameters, subharmonic oscillation occurs as shown in [7, Fig. 9].

Based on the *exact* sampled-data analysis by calculating the eigenvalues of $\Phi$, one pole is 0 and the second pole is -1.1. Based on (40), the second pole is exactly -1.1. For comparison, based on (42) as derived in [7], the second pole is -1.3. The error is due to the fact that the condition $T - d \ll RC$ is not met. The exact pole locus plot $S(\lambda, 0.4)$ based on (15) and the approximate one (19) are shown in Fig. 4. The intersections of the pole locus plot and the ramp slope ($m_a = 0$) are the two poles, -1.1 and 0. The approximate pole locus plot matches very well with the exact one, both accurately predicting the two poles.

The pole locus plot shows two types of information. First, it shows the resulting pole locations given a ramp slope $m_a$. For example, draw a horizontal line at $m_a$ in the pole locus plot, the intersections with the curve (15) (or (19)) are the pole locations. Second, given a desired pole location, the pole locus plot shows the required ramp slope to *assign* the pole to that particular location. For example, Fig. 4 shows that if the ramp slope is around 9500, the poles can be assigned at -0.5 and -0.2. □

**Example 2.** (*Pole assignment.*) Continued from Example 1, let $m_a = 9500$. Based on the *exact* sampled-data analysis by calculating the eigenvalues of $\Phi$, the poles are exactly -0.5 and -0.2 as predicted. □

**Example 3.** (*Determine the minimum ramp slope for stabilization.*) The ramp slope does not need to be as large as 9500 for stabilization (since only the poles being *within* the unit circle are required). Based on the *exact* sampled-data analysis by calculating the eigenvalues of $\Phi$, the minimum ramp slope for stabilization is 943.4. From (22), the minimum ramp slope for stabilization is also exactly 943.4. □

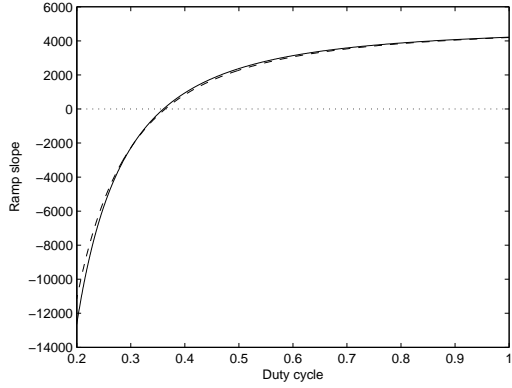

Figure 5: The S plot $S(-1, D)$ shows the required ramp slope to avoid PDB. The S plot also shows that, without a ramp ($m_a = 0$), PDB occurs at $D = 0.36$, (solid line for the exact S plot (22) and dashed line for the approximate S plot (24)).

**Example 4.** (*Determine the minimum ramp slope for a range of source voltage to maintain stability for line regulation.*) In Example 1, the source voltage is fixed at 5 V. Now suppose the converter is designed to operate in a range of source voltage $v_s \in [2, 10]$. With a fixed $d = 1.2$ $\mu$s and a fixed $v_o = 2$, then $D \in [0.2, 1]$, $T \in [1.2, 6]$ $\mu$s, and $\omega_s \in [1.047, 5.236]$ rad/s. Based on the exact sampled-data analysis, the converter is unstable for $D > 0.36$ (In Example 1, $D = 0.4 > 0.36$ and the converter is unstable).

The required ramp slope based on the exact S plot $S(-1, D)$ (22) and the approximate one based on (24) are shown in Fig. 5. The approximate one matches well with the exact one. The S plot shows exactly that the converter is unstable for $D > 0.36$ if no ramp is applied ($m_a = 0$). The S plot also shows that a ramp slope greater than 4217 is required to maintain stability for this operating range. □

**Example 5.** (*Determine the PDB point.*) Continued from Example 1 without a ramp, based on Fig. 5, PDB occurs at $D = 0.36$. With a fixed $d$ and a fixed $v_o = 2$, then PDB occurs at $T = 3.33$ $\mu$s and $v_s = 5.56$ V. With these parameters and based on the exact sampled-data analysis, the eigenvalues of $\Phi$ are exactly 0 and -1 as predicted. □

**Example 6.** (*Determine the maximum on-time.*) Continued from Example 1, by calculating the eigenvalues of $\Phi$, the maximum on-time $d$ to avoid subharmonic oscillation is 1.06 $\mu$s (with $T = 2.65$ $\mu$s to keep the duty cycle at 0.4). The maximum on-time based on (38), (39) and (41) are 0.84 $\mu$s, 0.8 $\mu$s, and 1.077 $\mu$s, respectively, and (41) gives the closest prediction. □

### 4.3 V-COTC Added With the Inductor Current as a Compensating Ramp

Here, the ramp $h(t)$ is not used, and $m_a = 0$. Instead, the inductor current is sensed through a resistance $R_i$ and added with the output voltage to compared with $v_c$ to determine the duty cycle as shown in Fig. 6.

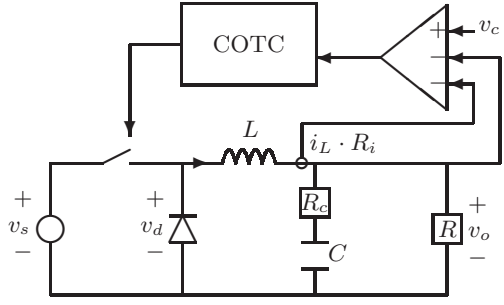

Figure 6: A buck converter under V-COTC with the inductor current as a compensating ramp.

The converter is turned on for a fixed time duration $d$. The converter changes from the OFF stage to the ON stage when $v_o + i_L R_i$ drops below $v_c$. One has $y = v_o + i_L R_i - v_c$ and

$$C = \rho[R_c, 1] + [R_i, 0] \qquad D = [0, -1]$$
$$CB_{11} = \frac{\rho R_c + R_i}{L} \qquad CA_1 B_{11} = \frac{\rho^2}{LC}(1 - \frac{R_c^2 C}{L} - \frac{R_c C R_i}{\rho L})$$

The boundary condition (27) becomes

$$\frac{d}{2} < \frac{(\rho R_c + R_i) C}{\rho^2 (1 - \frac{R_c^2 C}{L} - \frac{R_c C R_i}{\rho L})} \tag{44}$$

For $R_c(R_c + R_i)C \ll L$ and $R_c \ll R$, (44) is further simplified to

$$\frac{d}{2} < (R_c + R_i) C \tag{45}$$

which agrees with [9, Eq. 16]. Note that if the conditions $R_c(R_c + R_i)C \ll L$ and $R_c \ll R$ are not met, (44) will be more accurate than the boundary condition (45) as obtained in [9], and (45) is a special case of (44).

The pole location can be also expressed in a closed form. With some approximations for $R_c T \ll L$, (21) leads to

$$\lambda = 1 + \frac{2\rho T}{2(R_c + R_i)C + T - d}\left(\frac{T-d}{2RC} - 1 + \frac{R_i(T-d)}{2L}\right) \tag{46}$$

By setting $\lambda > -1$ in (46), the minimum of $R_i$ to avoid the subharmonic oscillation is

$$R_i > \frac{2d - 4R_c C - \frac{\rho T(T-d)}{RC}}{4C + \frac{\rho T(T-d)}{L}} \tag{47}$$

Based on some examples, (47) is more accurate than (44) and (45).

**Example 7.** (*Better prediction of minimum $R_i$.*) Continued from Example 1, by calculating the eigenvalues of $\Phi$, the minimum $R_i$ to avoid the subharmonic oscillation is 1.82 m$\Omega$. The minimum $R_i$ based on (44), (45) and (47) are 8.4 m$\Omega$, 10 m$\Omega$, and 3.4 m$\Omega$, respectively, and (47) gives the closest prediction. □

### 4.4 C-COTC

In C-COTC, one has $y = R_i i_L - v_c$ and

$$C = [R_i, 0] \qquad D = [0, -1]$$
$$CB_{11} = \frac{R_i}{L} \qquad CA_1 B_{11} = \frac{-\rho R_i R_c}{L^2} \qquad (48)$$

The boundary condition (26) to avoid SNB becomes

$$-\frac{Dv_s R_i}{2L}\left(\frac{d\rho R_c}{2L} + 1\right) < m_a \qquad (49)$$

The inequality is always satisfied even without the ramp ($m_a = 0$), and PDB does not occur, agreed with [8].

Also based on the *pole location*, one can prove that PDB does not occur in C-COTC as follows. With some approximations for $R_c T \ll L$, (21) leads to

$$\lambda = 1 - \frac{\rho T(T-d)}{2LC} \qquad (50)$$

Generally $T^2 \ll LC$, then $\lambda \approx 1$ and PDB (which requires $\lambda = -1$) does not occur. This sampled-data pole can be mapped to an equivalent continuous-time pole,

$$\frac{\rho(T-d)}{2LC} \qquad (51)$$

which will be shown to be more accurate than the pole $(2L/R - d)/2LC$ as derived in [8].

**Example 8.** (*Better prediction of the continuous-time pole.*) Consider a C-COTC buck converter from [8] with the following parameters: $v_s = 13.2$ V, $v_o = 3.3$ V, $T = 1.04$ µs, $d = 0.26$ µs, $R = 10$ Ω, $R_c = 4.5$ mΩ, $R_i = 150$ mΩ, $L = 3.1$ µH, and $C = 300$ µF. Here, $D = d/T = 0.25$.

Based on the *exact* sampled-data analysis by calculating the eigenvalues of $\Phi$, one pole is 0 and the second pole is 0.9995 (corresponding to a continuous-time pole at 473 Hz). Based on (51), the second continuous-time pole is 419 Hz. For comparison, based on [8], the second pole is $(2L/R - d)/2LC = 194$ Hz. This example shows that (51) gives a better prediction of the continuous-time pole.

The exact pole locus plot $S(\lambda, 0.25)$ based on (15) and the approximate one based on (19) are shown in Fig. 7. The approximate one matches very well with the exact one (except that the exact one has a finite ramp slope), both accurately predicting the two poles around 0 and 1. The pole locus plot also shows that if the ramp slope is small and negative, the two poles may be unstable (with the first pole smaller than -1 and the second pole greater than 1), which will be discussed in Example 9. □

## 5 Sampled-Data Analysis of SNB in the Buck Converter

Occurrence of SNB in COTC has not been reported. As discussed above, for V-COTC with a *negative* ramp slope, SNB may occur. In steady state,

$$y^0(T) - h(T) = Cx^0(T) + Du - h(T) = 0 \qquad (52)$$

which is an equation in terms of $T$. Generally, if this equation has multiple solutions of $T$, SNB may occur if a converter parameter varies. When SNB occurs, only a *single* solution exists, which means that the curve $Cx^0(T) + Du - h(T)$ has a flat slope (or equivalently, the curve $Cx^0(T) + Du$ is *tangential* to $h(T)$) when SNB occurs. An SNB is also called a tangent bifurcation.

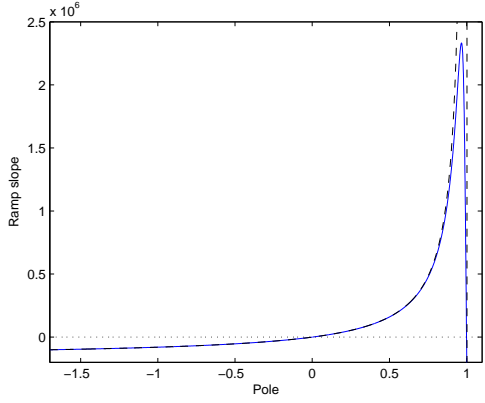

Figure 7: The pole locus plot $S(\lambda, 0.25)$ shows the poles around 0 and 1 without the ramp, (solid line for the exact expression (15) and dashed line for the approximate expression (19)).

Take the buck converter for example. The buck converter generally has $A_1 = A_2$, $B_{21} = 0$, and $B_{12} = B_{22}$. Using (6), and differentiate (52) with respective to $T$, one has the SNB boundary condition,

$$S(1, D) = C(I - e^{A_1 T})^{-2}(e^{A_1 T} - e^{A_1 T(1-D)})B_{11}v_s = m_a \tag{53}$$

One can derive or prove (53) in another way. Using (6), the boundary condition (18) directly leads to (53). It is interesting to know that the boundary conditions derived based on the steady-state analysis and based on the stability (eigenvalue) analysis are the same.

Based on the assumption that $RC$ and $\sqrt{LC}$ are much larger than $T$, matrix approximations such as $e^{A_1 T} \approx I + A_1 T + A_1^2 T^2/2$ and $(I + A_1 T)^{-1} \approx I - A_1 T$ can be applied. Then, the boundary condition (53) leads to

$$S(1, D) \approx (\frac{D}{T}CA_1^{-1}B_{11} - \frac{D^2}{2}CB_{11} + (\frac{2D^3 - D}{12})CA_1 B_{11} T)v_s = m_a \tag{54}$$

## 5.1 V-COTC: No SNB With a Positive Ramp Slope

Using (31) which results in $CA_1^{-1}B_{11} = -1$ for V-COTC, the boundary condition (54) becomes

$$-\frac{D}{T} - \frac{D^2 \rho R_c}{2L} + (\frac{2D^3 - D}{12})\frac{T\rho^2}{LC}(1 - \frac{R_c^2 C}{L}) \approx \frac{m_a}{v_s} \tag{55}$$

Generally, $R_c^2 C \ll L$ and $T^2 \ll 12LC$, and the left side of (55) is negative, whereas the right side is positive with a *positive* ramp slope. The boundary condition (55) cannot be met, and SNB does not occur in this situation.

## 5.2 C-COTC: No SNB With a Positive Ramp Slope

Using (48) which results in $CA_1^{-1}B_{11} = -R_i/R$, the boundary condition (54) becomes

$$-\frac{D}{TR} - \frac{D^2}{2L} - (\frac{2D^3 - D}{12})\frac{\rho R_c T}{L^2} \approx \frac{m_a}{v_s R_i} \tag{56}$$

Generally, $T^2 RR_c \ll 12L^2$, and the left side of (56) is negative, whereas the right side is positive with a *positive* ramp slope. The boundary condition (56) cannot be met, and SNB does not occur in this situation.

As discussed above, in C-COTC with $m_a = 0$, one pole is 0 and the second pole is very close to 1. The second pole may be shifted to be greater than 1, then SNB occurs. With a very small *negative* ramp slope, SNB may occur as shown in the next example.

**Example 9.** (*Pole assignment with a negative ramp slope.*) Continued from Example 8, based on (56), a negative ramp slope $m_a < -67445$ would result in an unstable pole greater than 1. Let $m_a = -100000$, which is $-0.42 s_f$, where $s_f := v_s D R_i / L$ is the off-time inductor current slope multiplied by $R_i$. Compared with $s_f$, this ramp slope is small. However, this ramp slope will result in SNB as shown below. Based on the exact sampled-data analysis by calculating the eigenvalues of $\Phi$, the poles are -1.675 and 1.0002 as predicted, also agreed with the pole locus plot shown in Fig. 7. Although C-COTC is generally believed to be stable even without the ramp, this example shows that a perturbation on the ramp slope to make it negative may make the converter unstable with occurrence of SNB. □

## 6 Harmonic Balance Analysis of PDB in the Buck Converter

Consider a buck converter power stage with a control-to-inductor-current ($D$-to-$i_L$) transfer function $G_{id}(s)$ and a control-to-output-voltage ($D$-to-$v_o$) transfer function $G_{vd}(s)$. In the converter, there is an ON switch and an OFF switch (sometimes substituted by a diode). Let the voltage across the OFF switch (or the diode) be $v_d$ (as shown in Fig. 2, for example). In the $T$-periodic mode, the waveform of $v_d(t)$ is a square wave with the high voltage at $v_s$ and the low voltage at 0, which can be represented by Fourier series (harmonics),

$$v_d(t) = \sum_{n=-\infty}^{\infty} c_n e^{jn\omega_s t} \quad \text{where} \quad c_n = \frac{v_s}{j 2 n \pi}(1 - e^{-jn\omega_s d}) \tag{57}$$

Similarly, in the $2T$-periodic mode (when PDB occurs), let two consecutive cycle periods be $T - \delta$ and $T + \delta$. Then, the switchings (from $S_1$ to $S_2$ and from $S_2$ to $S_1$) occurring within the two cycles are at $t = d$, $T - \delta$, $T - \delta + d$, and $2T$. The $2T$-periodic signal $v_d(t)$ can be also represented by Fourier series,

$$v_d(t) = \sum_{n=-\infty}^{\infty} c_n e^{\frac{jn\omega_s t}{2}} \quad \text{where} \quad c_n = (\frac{v_s}{-j 2 n \pi})(e^{-\frac{jn\omega_s d}{2}} - 1)(1 + (-1)^n e^{\frac{jn\omega_s \delta}{2}}) \tag{58}$$

In V-COTC, for example, $v_o$ is the feedback signal from the power stage (whereas in C-COTC, $i_L$ is the feedback signal from the power stage).

In the converter, some parts are linear (from $v_d$ to $y$) and some are nonlinear (from $y$ to $v_d$). Let the $v_d$-to-$v_o$ transfer function be $G_v(s)$. One has [15, p. 470]

$$G_v(s) = \frac{G_{vd}(s)}{v_s} = \frac{sR_c C + 1}{LC(1 + \frac{R_c}{R})s^2 + (\frac{L}{R} + R_c C)s + 1} \tag{59}$$

Let the compensator transfer function (from $v_o$ to $-y$ (negative sign due to negative feedback)) be $G_c(s)$. Let the total transfer function from $v_d$ to $-y$ be $G(s)$. Then, one has $G(s) = G_c(s)G_v(s) = G_c(s)G_{vd}(s)/v_s$. The gain $G(s)$ is proportional to the loop gain (based on the average analysis)

$$T(s) = \frac{G_c(s)G_{vd}(s)}{V_h} \tag{60}$$

by

$$G(s) = \frac{V_h}{v_s}T(s) \tag{61}$$

In the $s$-domain, $y = v_c + G_c(s)(v_c - v_o) = v_c + G_c(s)(v_c - G_v(s)v_d) = (1 + G_c(0))v_c - G(s)v_d$. Let **Re** denote taking the real part of a complex number. Then, from (57), the T-periodic solution $y^0(t)$ (at the output of the compensator) is

$$y^0(t) = (1 + G_c(0))v_c - \sum_{n=-\infty}^{\infty} c_n e^{jn\omega_s t} G(jn\omega_s) \tag{62}$$

$$= (1 + G_c(0))v_c - v_s DG(0) - 2\mathbf{Re}\sum_{n=1}^{\infty} c_n e^{jn\omega_s t} G(jn\omega_s) \tag{63}$$

The intersection of $h(t)$ with the T-periodic solution $y^0(t) = Cx^0(t) + Du$ determines the duty cycle and hence the waveform of $v_d(t)$. By "balancing" the equation $y^0(t) = h(t)$ (written in Fourier series form) at the switching instants, conditions for existence of periodic solutions and PDB can be derived.

## 6.1 PDB Boundary Conditions Based on Harmonic Balance Analysis

In the $2T$-periodic mode, one has two switching conditions

$$y^0(T - \delta) = h(T - \delta) \tag{64}$$
$$y^0(2T) = h(2T) \tag{65}$$

The basic idea to derive the PDB point is as follows. At the PDB point, a period-one mode and a period-two mode coalesce, and one has $\delta = 0$. Subtracting (64) from (65) and setting $\delta = 0$, one can obtain the PDB boundary condition

$$\frac{v_s}{2T}(\sum_{n=-\infty}^{\infty}(e^{-\frac{jn\omega_s d}{2}} - 1)(-1)^n G(\frac{jn\omega_s}{2})) = m_a \tag{66}$$

or equivalently,

$$\frac{v_s}{T}(\mathbf{Re}[\sum_{n=1}^{\infty}(e^{-\frac{jn\omega_s d}{2}} - 1)(-1)^n G(\frac{jn\omega_s}{2})]) = m_a \tag{67}$$

Since (66), (67) (based on the harmonic balance analysis), and (22) (based on the sampled-data analysis) are exact PDB conditions, they are all equivalent to each other. One has

$$S(-1, D) = \frac{v_s}{T}(\mathbf{Re}[\sum_{n=1}^{\infty}(e^{-\frac{jn\omega_s DT}{2}} - 1)(-1)^n G(\frac{jn\omega_s}{2})]) \tag{68}$$

which is another expression of the S plot in terms of signal harmonics.

Generally, $G_v(s)$, $G_c(s)$ and thus $G(s) = G_c(s)G_v(s)$ are low-pass filters. The summation in (67) can be approximated by the term that involves $G(s)$ with the smallest argument, $n = 1$ for example. Therefore, the S plot has an approximate form,

$$S(-1, D) \approx \frac{v_s}{T}(\mathbf{Re}[1 - (e^{-\frac{j\omega_s DT}{2}})G(\frac{j\omega_s}{2})]) \tag{69}$$

**Example 10.** (*Both harmonic balance analysis and sampled-data analysis produce the same plot.*) Continued from Example 4, with V-COTC, $y = v_o - v_c$. Then $G_c(s) = -1$ and $G(s) = -G_v(s)$. The plot of (66) in terms of $m_a$ is exactly the solid line in Fig. 5, indicating that (66) is equivalent to (22) (based on the sampled-data analysis). □

## 6.2 Harmonic Balance Analysis of PDB in Terms of the Loop Gain

Since the gain $G(s)$ is proportional to the loop gain $T(s) = G(s)v_s/V_h = G(s)v_s/m_aT$, (66) directly leads to the following theorem.

**Theorem 2** *Given a closed-loop COTC buck converter with a loop gain $T(s)$ as defined in (60), PDB occurs when*

$$\sum_{n=-\infty}^{\infty} (e^{-\frac{jn\omega_s d}{2}} - 1)(-1)^n T(\frac{jn\omega_s}{2}) = 2 \tag{70}$$

The condition (70) can be expressed in various forms, for example,

$$\mathbf{Re}[\sum_{n=1}^{\infty} (e^{-\frac{jn\omega_s d}{2}} - 1)(-1)^n T(\frac{jn\omega_s}{2})] = 1 \tag{71}$$

The summation in (71) can be also approximated by the term that involves $T(s)$ with the smallest argument. Therefore, (71) becomes

$$\mathbf{Re}[(1 - e^{-\frac{j\omega_s d}{2}})T(\frac{j\omega_s}{2})] \approx 1 \tag{72}$$

It should be noted that (72) is only an approximate condition, and the exact condition is (70).

**The "L1 plot" in the real domain.**
Note that the boundary condition (71) is a function of $d$, $\omega_s$, and the loop gain $T(s)$, where $T(s)$ is further a function of $v_s$, $m_a$, the power stage and compensator parameters. Define an L1 plot, which is a *real* function, as

$$L_1(\omega_s) := \sum_{n=-\infty}^{\infty} (e^{-\frac{jn\omega_s d}{2}} - 1)(-1)^n T(\frac{jn\omega_s}{2}) \tag{73}$$

Then, SNB occurs when

$$L_1(\omega_s) = 2 \tag{74}$$

Note that $\omega_s = 2\pi/T = 2\pi D/d$, which is linearly proportional to $D$ given a fixed value of $d$ in COTC, $L_1(\omega_s)$ can be represented as a function of $D$ instead of $\omega_s$.

**The "L2 plot" in the real domain.**
Since in some situations, $m_a = 0$ when no ramp compensation is used, then $T(s) = G(s)v_s/V_h$ becomes infinite. In such a situation, define an L2 plot, which is also a *real* function, as

$$L_2(\omega_s) := \sum_{n=-\infty}^{\infty} (e^{-\frac{jn\omega_s d}{2}} - 1)(-1)^n G(\frac{jn\omega_s}{2}) \tag{75}$$

Then, from (66), SNB occurs when

$$L_2(\omega_s) = \frac{2Tm_a}{v_s} \quad (76)$$

**Example 11.** (*Fig. 5 is also an L2 plot*) Continued from Example 4, since $\omega_s = 2\pi/T = 2\pi D/d$ is linearly proportional to $D$, Fig. 5 is also an L2 plot with the horizontal axis scaled by $\omega_s = 2\pi/T = 2\pi D/d$. □

**The "H plot": a Nyquist-like plot in the complex plane.**
Let

$$H(\omega_s) := \sum_{n=1}^{\infty} (e^{-\frac{jn\omega_s d}{2}} - 1)(-1)^n G(\frac{jn\omega_s}{2}) \quad (77)$$

Then, PDB occurs when

$$\mathbf{Re}[H(\omega_s)] = \frac{Tm_a}{v_s} \quad (78)$$

For designation purpose, $H(\omega_s)$ is called an H plot because it is similar to the Nyquist plot in the complex plane. Given a desired range of $\omega_s = 2\pi/T = 2\pi D/d$, one can plot $H(\omega_s)$ according to (77) to determine whether PDB occurs in this range of $\omega_s$.

### 6.3 V-COTC

As shown below, the harmonic balance analysis also leads to the same conditions derived by the sampled-data analysis. In V-COTC, $G_c(s) = -1$, and $G(s) = G_c(s)G_v(s) = -G_v(s)$. Using (59) and the facts that $\omega_s d/2 = \pi D$ and, for $0 < D < 1$,

$$\sum_{k=1}^{\infty} \frac{(1 - \cos(\pi k D))(-1)^k}{k^2} = -\frac{\pi^2 D^2}{4} \quad (79)$$

$$\sum_{k=1}^{\infty} \frac{\sin(\pi k D)(-1)^k}{k} = -\frac{\pi D}{2} \quad (80)$$

the boundary condition (67) becomes

$$\frac{D}{2LC}(\frac{d}{2} - R_c C) < \frac{m_a}{v_s} \quad (81)$$

which is exactly (35), but expressed in a different form. It is interesting to note that both the sampled-data analysis and the harmonic balance analysis lead to the same boundary condition, providing convincing evidence about the accuracy of the derived boundary condition.

### 6.4 C-COTC

In C-COTC, the feedback signal from the power stage is $i_L$. Similar to (59), the $v_d$-to-$i_L$ transfer function is [15, p. 470]

$$G_i(s) := \frac{G_{id}(s)}{v_s} = \frac{(1 + \frac{R_c}{R})Cs + \frac{1}{R}}{LC(1 + \frac{R_c}{R})s^2 + (\frac{L}{R} + R_c C)s + 1} \quad (82)$$

Since no extra compensator (except the compensating ramp $h(t)$) is added in the current loop, $G_c(s) = -1$, and the gain $G(s) = G_c(s)G_i(s) = -G_i(s)$. Since the expression of $G(s)$ is derived, the rest of analysis is similar to the case for V-COTC.

# 7  Harmonic Balance Analysis of SNB in the Buck Converter

Using (57) and (62), in steady state,

$$y^0(T) = (1 + G_c(0))v_c - \sum_{n=-\infty}^{\infty} \frac{v_s}{j2n\pi}(e^{\frac{-j2n\pi d}{T}} - 1)G(\frac{j2n\pi}{T}) \tag{83}$$

As discussed above, the curve $y^0(T) - h(T)$ has a flat slope (or equivalently, the curve $y^0(T)$ is tangential to $h(T)$) when SNB occurs. Differentiating (83) with respective to $T$, one has the SNB boundary condition,

$$\sum_{n=-\infty}^{\infty} \frac{d}{T^2} e^{-jn\omega_s d} G(jn\omega_s) + (1 - e^{-jn\omega_s d}) \left.\frac{\partial G(s)}{\partial s}\right|_{s=jn\omega_s} = \frac{m_a}{v_s} \tag{84}$$

# 8  Conclusion

Design-oriented bifurcation boundary conditions are derived for general switching DC-DC converters under COTC. The derived boundary conditions define the instability boundaries in the converter parameter space, and are therefore useful for a designer to design a converter to avoid the occurrence of the bifurcations or instabilities. Two bifurcations commonly seen in DC-DC converters are analyzed. They are PDB and SNB. Two COTC schemes, V-COTC and C-COTC, are also are analyzed based on a unified model. The derived boundary conditions are more general and accurate than those recently obtained in [7, 8, 9]. The boundary conditions *recently* obtained in [7, 8, 9] become special cases under the general modeling approach presented in this paper. Since the analysis is based on the *general* block diagram model shown in Fig. 1, once a converter (of any system dimension under various control schemes) is expressed in terms of the block diagram model, the boundary conditions can be readily obtained.

Both *exact* sampled-data analysis and harmonic balance analysis are applied. Different analyses give different perspectives on the system dynamics and complement each other. Under the sampled-data analysis, the boundary conditions are expressed in terms of signal slopes and the ramp slope. Under the harmonic balance analysis, the boundary conditions are expressed in terms of signal harmonics. Although occurrence of SNB in the converter under COTC is rare, it is shown that a small perturbation on the ramp to make the ramp slope negative may cause the occurrence of SNB.

Many new design-oriented plots are proposed. The S plot predicts PDB and SNB instabilities and the required ramp slope to avoid them. A pole locus plot, a special case of the S plot, predicts the real pole location and shows the required ramp slope for pole assignment. The S plot can be expressed in terms of matrices or signal harmonics. The S plot in terms of signal harmonics also leads to other design-oriented plots, such as the L1 and L2 plots in the real domain, and the H plot in the complex plane, to facilitate stability analysis and design.